\documentclass[11pt,reprint,onecolumn,notitlepage,longbibliography,superscriptaddress]{revtex4-1}

\usepackage{graphicx}
\usepackage{amsmath} 
\usepackage{mathtools}
\usepackage{amsfonts}
\usepackage{bbm}
\usepackage{braket}
\usepackage{caption}
\usepackage{graphicx}
\usepackage{float}
\usepackage{caption}
\usepackage{subcaption}
\usepackage{url}
\usepackage{color}
\usepackage{qcircuit}
\usepackage{caption}
\usepackage{kbordermatrix}

\usepackage{setspace}
\usepackage[margin=1cm]{caption}

\begin{document}
\title{Quantum Computing for Atomic and Molecular Resonances}
\author{Teng Bian}
\affiliation{Department of Chemistry, Purdue University, West Lafayette, IN, 47907 USA}
\affiliation{Department of Physics and Astronomy, Purdue University, West Lafayette, IN, 47907 USA}
\affiliation{Purdue Quantum Science and Engineering Institute, Purdue University, West Lafayette, IN, 47907 USA}

\author{Sabre Kais}
\email{kais@purdue.edu}
\affiliation{Department of Chemistry, Purdue University, West Lafayette, IN, 47907 USA}
\affiliation{Department of Physics and Astronomy, Purdue University, West Lafayette, IN, 47907 USA}
\affiliation{Purdue Quantum Science and Engineering Institute, Purdue University, West Lafayette, IN, 47907 USA}

\begin{abstract}
The complex-scaling method can be used to calculate molecular resonances within the Born-Oppenheimer approximation, assuming the electronic coordinates are dilated independently of the nuclear coordinates. With this method, one will calculate the complex energy of a non-Hermitian Hamiltonian, whose real part is associated with the resonance position and the imaginary part is the inverse of the lifetime. In this study, we propose techniques to simulate resonances on a quantum computer. First, we transformed the scaled molecular Hamiltonian to second-quantization and then used the  Jordan-Wigner transformation to transform the scaled Hamiltonian to the qubit space. To obtain the complex eigenvalues, we introduce the Direct Measurement method, which is applied to obtain the resonances of a simple one-dimensional model potential that exhibits pre-dissociating resonances analogous to those found in diatomic molecules.  Finally, we applied the method to simulate the resonances of the H$_2^-$ molecule.  
Numerical results from the IBM Qiskit simulators and IBM quantum computers verify our techniques. \\
\\
\noindent
[Keywords]: Quantum computing, molecular systems, resonances, Direct Measurement method

\end{abstract}
\maketitle

\section{Introduction}

Resonances are intermediate or quasi-stationary states that exist during unique atomic processes such as: when an excited atom autoionizes, an excited molecule disassociates unimolecularly, or a molecule attracts an electron and then the ion disassociates into stable ionic and neutral subsystems\cite{reinhardt1982complex}. The characteristics of resonances, such as energy and lifetime, can be revealed by experiments or predicted by theory. One theoretical method to compute properties associated with such resonances is called the complex-scaling method, developed by \cite{aguilar1971class, balslev1971spectral, simon1972quadratic, simon1973resonances, van1974complex, moiseyev1998quantum}. This method is based on the Balslev-Combes theorem, which is valid for dilation-analytic potentials and can be extended for non-dilation-analytic potential energies \cite{moiseyev1979autoionizing}. Additionally, several variants have been developed to study problems like Stark resonances\cite{emmanouilidou2000scattering, orimo2018implementation, jagau2018coupled} induced by an external electric field. The real space extension of this method uses standard quantum chemistry packages and stabilization graphs\cite{haritan2017calculation}. Its main applications are to study the decay of metastable states existing above the ionization threshold of Li centre in open-shell systems LiHe \cite{landau2020ab}, in the computation of transition amplitudes among metastable states\cite{bhattacharya2020ab}, and in explaining Autler-Townes splitting of spectral lines \cite{pick2019ab}.

The complex-scaling method usually requires a large basis set to predict resonances with good accuracy. For example, the Helium $^1S$ resonance uses 32 Hylleraas type functions for basis construction \cite{moiseyev1978complex}, and the $H_2^-$ $^2\Sigma_u^+(\sigma_g^2\sigma_u)$ resonance takes a total of $38$ constructed Gaussian atomic bases \cite{moiseyev1979autoionizing}. Computational overhead will become overwhelming if more basis functions need to be considered, like when simulating larger molecular systems, or requiring higher accuracy. Moreover, dimensional scaling and large-order dimensional perturbation theory have been applied for complex eigenvalues using the complex-scaling method \cite{kais1993dimensional, germann1993large}. Like for bound states\cite{kais_book,xia2018quantum,bian2019quantum,daskin2019context,xia2017electronic}, quantum computing algorithms can overcome the above computational limitation problem for resonances. However, most algorithms cannot be directly adapted to resonance calculation with the complex-scaling method because the complex-rotated Hamiltonian is non-Hermitian. For example, the propagator $e^{-iH(\mathbf{r}e^{i\theta})t}$ in the conventional phase estimation algorithm (PEA) with trotterization \cite{lloyd1996universal} will be non-unitary and it cannot be implemented in a quantum circuit directly. In this way, a quantum algorithm for resonance calculation that can work with non-Hermitian Hamiltonians is needed. Daskin et al. \cite{daskin2014universal} proposed a circuit design that can solve complex eigenvalues of a general non-unitary matrix. The method applies the matrix rows to an input state one by one and estimates complex eigenvalues via an iterative PEA process. However, for molecular Hamiltonians, the gate complexity of this general design is exponential in system size. In our previous publication\cite{bian2019quantum}, we briefly mentioned that our Direct Measurement method can solve complex eigenvalues of non-Hermitian Hamiltonians with polynomial gates. This study extends the Direct Measurement method and applies it to simple molecular systems as benchmark tests to obtain resonance properties. In particular, we will use IBM's Qiskit \cite{Qiskit} simulators and their quantum computers to calculate these resonances.

In the following sections, we first show how to obtain the complex-scaled Hamiltonian for molecular systems and transform it into Pauli operator form. Then, we introduce the Direct Measurement method that can derive the Hamiltonian's complex eigenvalues. Finally, we apply this method to do resonance calculation for a simple model system and a benchmark test system $H_2^-$, using simulators and IBM quantum computers.

\section{Complex Scaled Hamiltonian}
This section presents the steps needed to convert the 
complex-rotated Hamiltonian to a  suitable form that can be simulated on a quantum computer. In the Born-Oppenheimer approximation,  the electronic Hamiltonian of a molecular system can be written as a sum of electronic kinetic energy and potential energy of the form,
\begin{align}
\begin{split}
    &H(\mathbf{r}) = T(\mathbf{r}) + V(\mathbf{r}),\\
    &T(\mathbf{r}) = \sum_i -\frac{1}{2}\nabla_i^2,\\
    &V(\mathbf{r}) = \sum_{i,j} \frac{1}{|\mathbf{r_i} - \mathbf{r_j}|} + \sum_{i, \sigma}\frac{Z_\sigma}{|\mathbf{r}_i - \mathbf{R}_\sigma|},
\end{split}
\label{eq:Ham_eqn}
\end{align}
where $Z_\sigma$ is the $\sigma_{th}$ nucleus' charge, $\mathbf{R}_\sigma$ is the $\sigma_{th}$ nucleus' position, and $\mathbf{r}_i$, $\mathbf{r}_j$ represents the $i_{th}$, $j_{th}$ electron's position.  The complex scaling method is applied to the study of molecular resonances within the framework of Born-Oppenheimer approximation.  Following Moiseyev et al.\cite{moiseyev2011non}, the electronic coordinates are dilated independently of the nuclear coordinates. 
Given such a Hamiltonian $H(\mathbf{r})$ in Eq. (\ref{eq:Ham_eqn}), where $\mathbf{r}$ represents electrons' coordinates, the complex-scaling method rotates $\mathbf{r}$ into the complex plane by $\theta$,  $\mathbf{r} \xrightarrow{} \mathbf{r}e^{i\theta}$. Thus the Hamiltonian becomes $H(\mathbf{r} e^{i\theta})$. After a complex rotation by $\theta$, each electron's position $\mathbf{r}$ becomes $\mathbf{r}/\eta$, where $\eta=e^{-i\theta}$ and thus the new Hamiltonian from Eq. (\ref{eq:Ham_eqn}) becomes
\begin{align}
    &H_\theta = T(\mathbf{r}/\eta) + V(\mathbf{r}/\eta),\\
    &T(\mathbf{r}/\eta) = \eta^2 \sum_i - \frac{1}{2}\nabla_i^2,\\
    &V(\mathbf{r}/\eta) = \eta \sum_{i \neq j} \frac{1}{|\mathbf{r}_i - \mathbf{r}_j|} + \eta \sum_{i, \sigma}\frac{Z_\sigma}{|\mathbf{r}_i - \eta \mathbf{R}_\sigma|}.
\end{align}

It is shown that the system's resonance state's energy $E$ and width $\Gamma = \frac{1}{\tau}$, where $\tau$ is the life time, are related to the corresponding complex eigenvalue of $H(\mathbf{r} e^{i\theta})$,\cite{balslev1971spectral, moiseyev1978resonance}
\begin{align}
    E_\theta = E - \frac{i}{2}\Gamma.
    \label{eq:energy_theta}
\end{align}
When doing exact calculation in an infinite basis limit, $E_\theta$ in Eq.(\ref{eq:energy_theta}) is not a function of $\theta$. However, there would be dependence in reality because only a truncated basis set is always used in practice. The best resonance estimate is when the complex energy $E_{\theta}$ pauses or slows down in its trajectory \cite{doolen1975procedure, moiseyev1978resonance} in the $(E_{\theta},\theta)$ plane or $\frac{d E_{\theta}}{d \theta} = 0$. In this way, $E$ and $\Gamma$ can be obtained by solving the new Hamiltonian's eigenvalues for $\theta$ trajectories, and looking for the pause. A scaling parameter $\alpha$ is commonly used in the complex rotation process to locate better resonances, which makes $\eta = \alpha e^{-i\theta}$. We refer the readers to the book on non-Hermitian quantum mechanics by Moiseyev for more details and method applications \cite{moiseyev2011non}.

After choosing a proper orthogonal basis set $\{\psi_i(\mathbf{r})\}$, the Hamiltonian can be converted into a second quantization form,
\begin{align}
    &H_\theta = \sum_{i,j}h_{ij} a_i^{\dagger}a_j + \frac{1}{2}\sum_{i,j,k,l} h_{ijkl} a_i^\dagger a_j^\dagger a_k a_l,
    \label{ham_second_quantization}
\end{align}
In the equation, $a_i^\dagger$ and $a_i$ are fermionic creation and annihilation operators. The coefficients $h_{ij}$, $h_{ijkl}$ can be calculated by
\begin{align}
\begin{split}
    &h_{ij} = \int \psi_i^*(\mathbf{r}) (-\eta^2\frac{1}{2}\nabla_i^2 + \eta \sum_{\sigma}\frac{Z_\sigma}{|\mathbf{r} - \eta \mathbf{R}_\sigma|}) \psi_j(\mathbf{r}), \\
    &h_{ijkl} = \int \psi_i^*(\mathbf{r_1}) \psi_j^*(\mathbf{r_2})  \frac{\eta}{|\mathbf{r}_1 - \mathbf{r}_2|} \psi_k(\mathbf{r_2}) \psi_l(\mathbf{r_1}).
\end{split}
\end{align}
With the Jordan-Wigner transformation\cite{seeley2012bravyi},
\begin{align}
\begin{split}
    a_j^\dagger = \frac{1}{2}(X_j - iY_j)\otimes Z_{j-1}^{\rightarrow},\\
    a_j = \frac{1}{2}(X_j + iY_j)\otimes Z_{j-1}^{\rightarrow},
\end{split}
\label{jordan_wigner}
\end{align}
in which $X, Y$ and $Z$ are the Pauli operators, and
\begin{align}
    Z_{j-1}^{\rightarrow} = Z_{j-1} \otimes Z_{j-2} \otimes Z_{0},
\end{align}
the Hamiltonian in Eq.(\ref{ham_second_quantization}) will be further transformed into Pauli operators as 
\begin{align}
    H_\theta = \sum_{i=0}^{L-1} c_i P_i.
    \label{hamiltonian_tensor_product}
\end{align}
In the summation, $c_i$ represents a complex coefficient, and $P_i$ represents a $k$-local tensor product of Pauli operators, where $k \leq n$ and $n$ is the size of the basis set. Alternatively, the Bravyi-Kitaev transformation \cite{seeley2012bravyi} or parity transformation can also be used in the final step for obtaining the Hamiltonian in the qubit space.\\

The above process is the same as the conventional Hamiltonian derivation in quantum computing for electronic structure calculations of bound states\cite{kais_book, o2016scalable, aspuru2005simulated, cao2019quantum, mcclean2020openfermion}. Here for resonance calculations, to make the Hamiltonian more compatible with the Direct Measurement method, we rewrite Eq.(\ref{hamiltonian_tensor_product}) as
\begin{align}
    & H_\theta = \sum_{i=0}^{2^{n_a}-1} \beta_i V_i,
    \label{hamiltonian_tensor_product_direct_measurement}
\end{align}
where $n_a = \lceil \log_2 L \rceil$. The coefficient $\beta_i$ and the operator $V_i$ are determined in the following ways,
\begin{align}
\begin{split}
    \beta_i = |c_i|, V_i = \frac{c_i}{|c_i|} P_i &\text{,\quad when $i < L$},\\
    \beta_i = 0, V_i = I &\text{,\quad when $i \geq L$}.
\end{split}
\end{align}

\section{Direct Measurement Method}
The Direct Measurement method is inspired by the direct application of the Phase Estimation Algorithm \cite{daskin2018direct} as briefly discussed in our previous publication\cite{ bian2019quantum}. Here the basic idea is to apply the complex-rotated Hamiltonian to the state of the molecular system and obtain the complex energy information from the output state. Since the original non-Hermitian  Hamiltonian cannot be directly implemented in a quantum circuit, this Direct Measurement method embeds it into a larger dimensional unitary operator.\\

\begin{figure}
    \centering
    \includegraphics{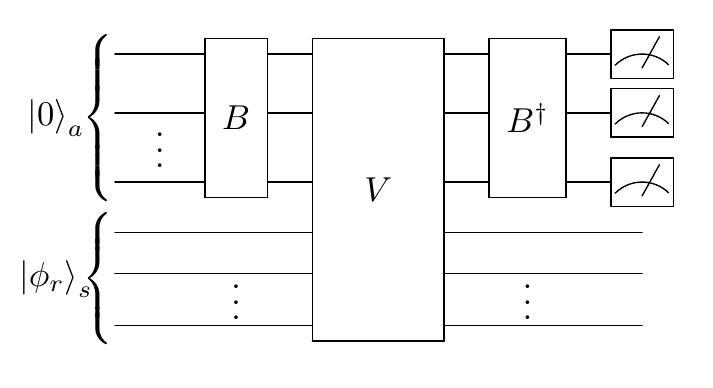}
    \caption{The quantum circuit for Direct Measurement method. $B$ and $V$ gates are constructed based on the coefficients and operators in Eq.(\ref{hamiltonian_tensor_product_direct_measurement}). The system qubits' state and ancilla qubits' state are initialized as $\ket{0}_a$ and $\ket{\phi_r}_s$ respectively.}
    \label{fig:direct_measurement_circuit}
\end{figure}

Assuming $n$ spin orbitals need to be considered for the system, the Direct Measurement method requires $n_s = n$ qubits to prepare the state of the model system  $\ket{\phi_r}_s$ and an extra $n_a$ ancilla qubits to  enlarge the non-Hermitian Hamiltonian to be a unitary operator. The quantum circuit is shown in FIG. \ref{fig:direct_measurement_circuit}.

\noindent
The $B$ and $V$ gates in the circuit are designed to have the following properties
\begin{align}
    & B\ket{0}_a = \sum_{i=0}^{2^{n_a}-1} \sqrt{\frac{\beta_i}{A}} \ket{i}_a, A = \sum_{i=0}^{2^{n_a}-1} \beta_i\\
    & V\ket{i}_a \ket{\phi_r}_s = \ket{i}_a V_i \ket{\phi_r}_s,
\end{align}
which means $B$ transforms the initial ancilla qubits' state to a vector of coefficients, and $V$ applies all $V_i$ on system qubits based on ancilla qubits' states. One construction choice for $B$ could be implementing the unitary operator
\begin{align}
    B = 2(\sum_{i=0}^{2^{n_a}-1} \sqrt{\frac{\beta_i}{A}} \ket{i}_a)(\sum_{i=0}^{2^{n_a}-1} \sqrt{\frac{\beta_i}{A}} \bra{i}_a)  - I.
    \label{eq:bgate_construction}
\end{align}
As for $V$, a series of multi-controlled $V_i$ gates will do the work. If $\ket{\phi_r}_s$ is chosen as an eigenstate and we apply the whole circuit of $B$, $V$, and $B^\dagger$
\begin{align}
    U_r = (B^{\dagger} \otimes I^{\otimes n_s}) V  (B \otimes I^{\otimes n_s}),
\end{align}
on it, the output state will be
\begin{equation}
    U_r \ket{0}_a \ket{\phi}_s = \frac{Ee^{i\varphi}}{A} \ket{0}_a \ket{\phi}_s + \ket{\Phi^\bot},
    \label{eq:output_state}
\end{equation}
where $E e^{i\varphi}$ ($E \ge 0$) is the corresponding eigenvalue and $\ket{\Phi^{\perp}}$ is a state whose ancilla qubits' state is perpendicular to $\ket{0}_a$. Then we can derive $E$ by measuring the output state. To obtain the phase $\varphi$, we apply a similar circuit for $H'_\theta = xI^{\otimes n} + H_\theta$, where $x$ is a selected real number, and perform the measurements. The calculation details are found in Appendix C.

\section{Quantum simulation of  resonances in a simple model system}
\begin{figure}
    \centering
    \includegraphics[width=0.6\linewidth]{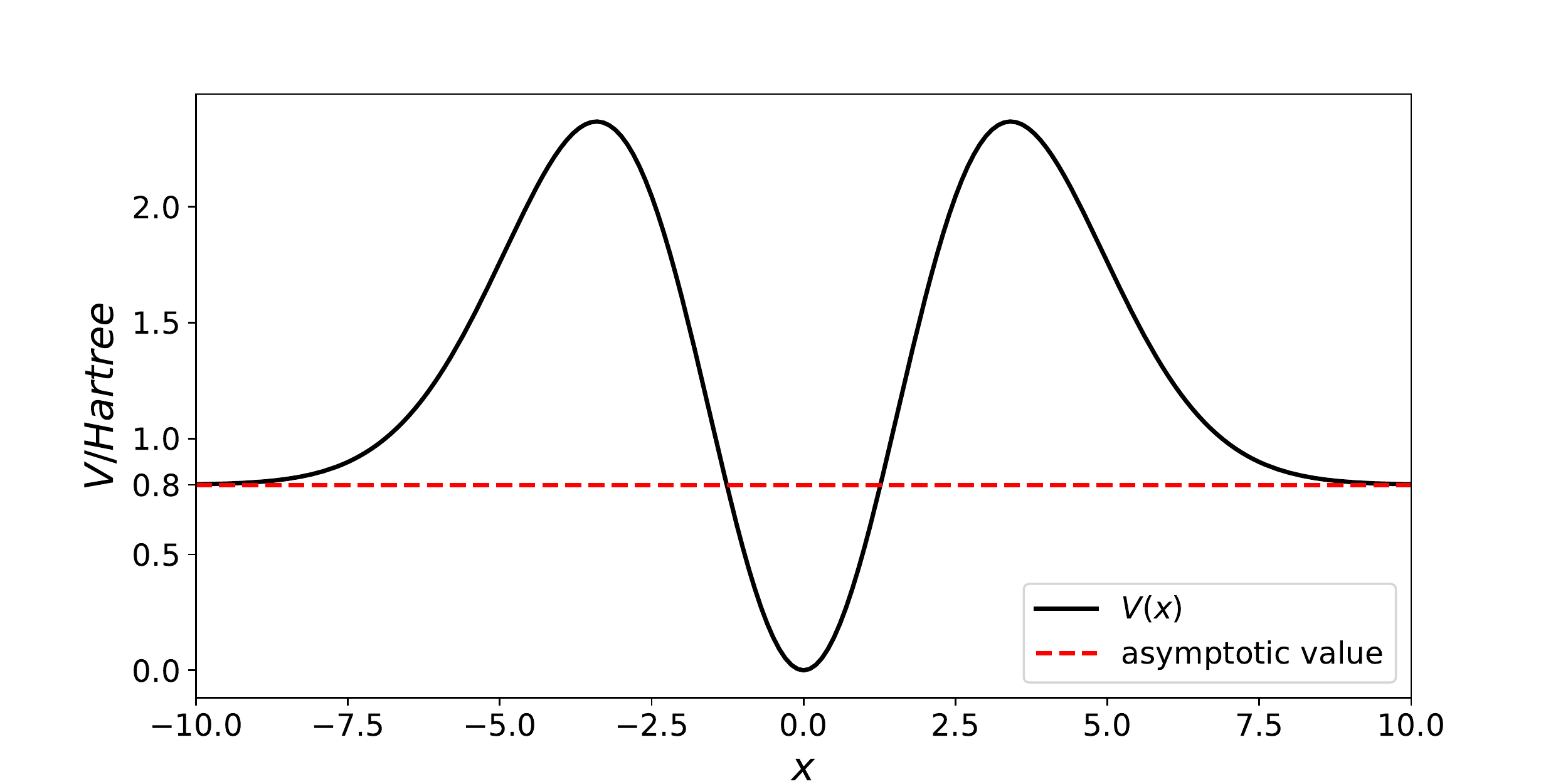}
    \caption{The one-dimensional potential $V(x) = (\frac{1}{2}x^2 - J) e^{-\lambda x^2} + J$, where $\lambda=0.1$, $J = 0.8$.}
    \label{fig:potential}
\end{figure}

In this section, we calculate the resonance properties of a model system using the Direct Measurement method. This system is the following one-dimensional potential\cite{moiseyev1978resonance}
\begin{align}
    V(x) = (\frac{1}{2}x^2 - J) e^{-\lambda x^2} + J,
\end{align}
and parameters are chosen as $\lambda=0.1$, $J = 0.8$. The potential plot is in FIG. \ref{fig:potential}.
This potential is used to model some resonance phenomena in diatomic molecules. We only consider one electron under this potential. The original Hamiltonian and the complex-rotated Hamiltonian can be written as
\begin{align}
    H &= -\frac{\nabla_x^2}{2} + V(x),\\
    H_\theta &= -\eta^2 \frac{\nabla_x^2}{2} + V(\eta x)
\end{align}
To make the setting consistent with the original literature, $\eta$ is chosen to be $e^{-i\theta}$ and the scaling parameter $\alpha$ is embedded in $n$ Gaussian basis functions
\begin{align}
    &\chi_k(\alpha) = \exp(-\alpha_k x^2),\\
    &\alpha_k = \alpha (0.45)^k, k = 0,1,..,n-1.
\end{align}
The $\{\chi_k(\alpha)\}$ basis set is not orthogonal, so we apply Gram-Schmidt process and iteratively construct an orthogonal basis set $\{ \psi_i \}$ as follows:
\begin{align}
    \gamma_k &= \chi_k - \sum_{i=0}^{k-1} \braket{\chi_k|\psi_i}\psi_i,\\
    \psi_i& = \frac{\gamma_k}{||\gamma_k||} = \frac{\gamma_k}{\sqrt{\braket{\gamma_k|\gamma_k}}}.
\end{align} 
Since there is only one electron, we do not consider spin interactions. This $\{ \psi_i \}$ basis set is used in the second quantization step to get the final Hamiltonian in Pauli matrix form. The resonance eigenvalue found in \cite{moiseyev1978resonance} with $n = 10$ basis functions is $E_\theta = 2.124 - 0.019i \quad$ Hartree. We will try to get the same resonance by applying the Direct Measurement method using the Qiskit package. The Qiskit package supports different backends, including a statevector simulator that executes ideal circuits, a QASM simulator that provides noisy gate simulation, and various quantum computers. In the following section, we show the results when the basis function number is $n = 5$ and $n = 2$. In particular, the former $n=5$ case shows how $\theta$ trajectories locate the best resonance estimate, and the latter $n=2$ cases show how to further simplify the quantum circuit for the Direct Measurement method and run it on IBM quantum computers.\\
\begin{table}[hbt!]
    \centering
    \begin{tabular}{|c|c|c|c|c|c|}
    \hline
        Case Name  & \#Basis functions & \#Total Qubits & \#System Qubits & \#Ancilla Qubits & \#Gates\\
    \hline
    \hline
        C1 & 5 &10 & 5 & 5 &  $\sim 10^6$\\
    \hline
        C2 & 2 & 5 & 2 & 3 & $\sim$ 800 \\
    \hline
        C3 & 2 & 4 & 2 & 2 & $\sim$ 200 \\
    \hline
        C4 & 2 & 3 & 2 & 1 & $\sim$ 10 \\
    \hline
    \end{tabular}
    \caption{The number of qubits and estimated gates in different cases when the Direct Measurement method is used to calculate resonance properties for the model system. The estimation for gate numbers is based on the QASM simulator and IBM machines.}
    \label{tab:qubits_gates}
\end{table}

C1 in the Table \ref{tab:qubits_gates} is our primary example where we follow the above steps in section II and III for $n=5$. An example of the complex-rotated Hamiltonian is shown in Appendix A. FIG. \ref{fig:direct_measurement_result} shows a sweep of scaling parameters $\alpha$, for statevector simulations of $\theta$ trajectories.
\begin{figure}
    \centering
    \includegraphics[width=0.7\linewidth]{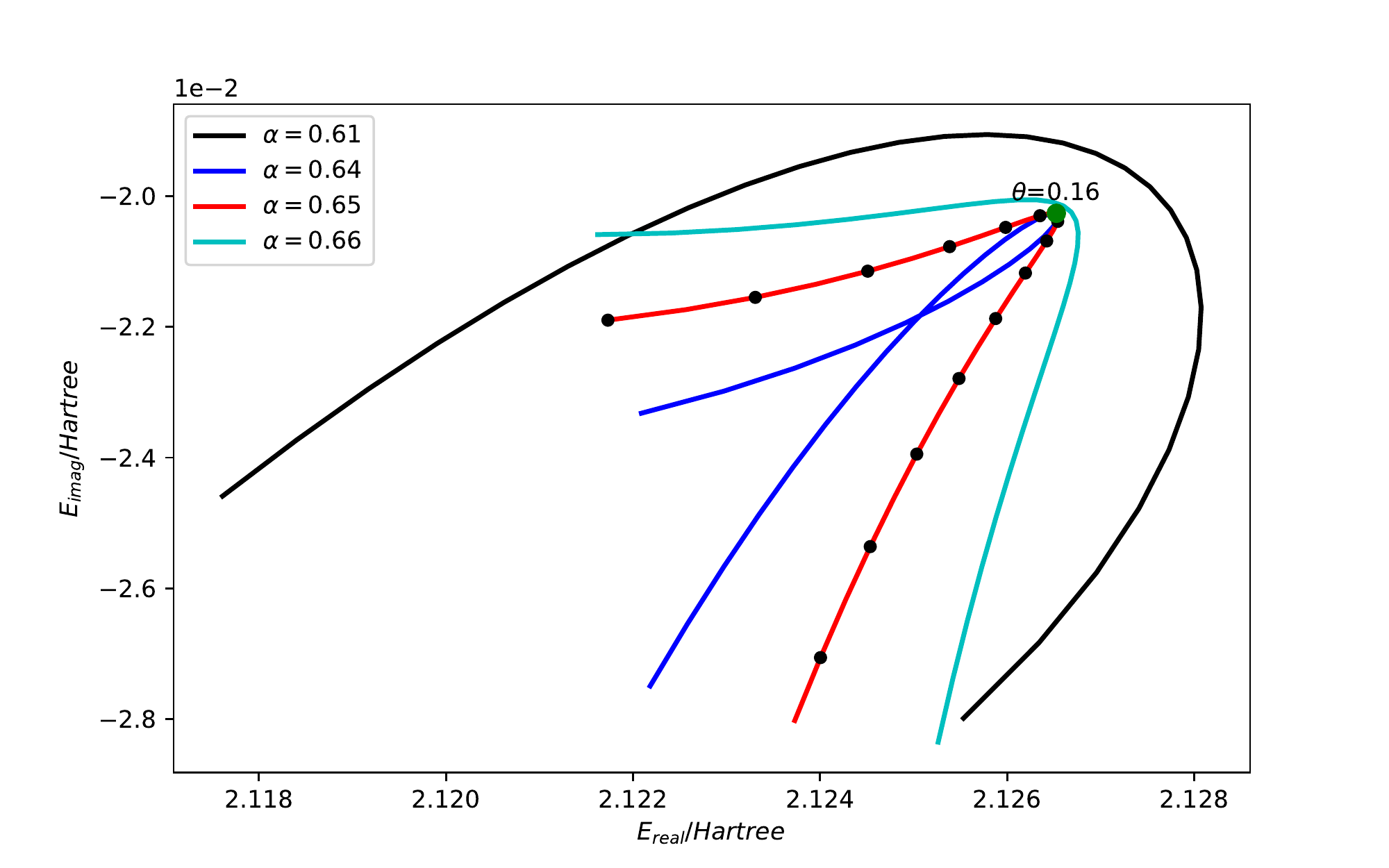}
    \caption{Trajectories of a complex eigenvalue on the rotation angle $\theta$ for fixed $n=5$ and various $\alpha$, calculated by Qiskit statevector simulator. $\theta$ ranges from 0.1 to 0.24 with a step of 0.01. The green point shows the best estimation of resonance energy, which is $E = 2.1265 - 0.0203i$ Hartree, occurs at $\alpha = 0.65, \theta = 0.160$. The input state for the Direct Measurement method is obtained by directly diagonalizing the complex-rotated Hamiltonian matrix.}
    \label{fig:direct_measurement_result}
\end{figure}
\noindent
Most trajectories pause around the point, $E_\theta = 2.1265 - 0.0203i$ Hartree,
when $\alpha = 0.65$, $\theta = 0.160$. Based on Eq.(\ref{eq:energy_theta}), this indicates the resonance energy and width are $E = 2.1265$ Hartree, $\Gamma = 0.0406$ Hartree,
 close to the resonance energy from \cite{moiseyev1978resonance}. The IBM quantum computer cannot perform the method due to a large number of standard gates in the circuit. Instead, we used the QASM simulator for $4*10^4$ shots and obtained the system's resonance energy at $\alpha = 0.65$, $\theta = 0.160$, $E_\theta = 2.1005 - 0.3862i $ Hartree. This result has an error of around $0.3$ Hartree but can be augmented by more sample measurements.  \\

When taking $n = 2$ for the basis function, we are not able to locate the best resonance estimate, see FIG. \ref{fig:direct_measurement_result}, based on direct diagonalization. So, we only use the Direct Measurement method to calculate the complex eigenenergy when $\alpha = 0.65$ and $\theta = 0.160$, where the best location is at $n=5$. We run the Direct Measurement method using simulators first and then try to reduce the number of ancilla qubits to make the resulting circuit short enough to be executed in the IBM quantum computers.\\

\begin{figure}
    \centering
    \includegraphics{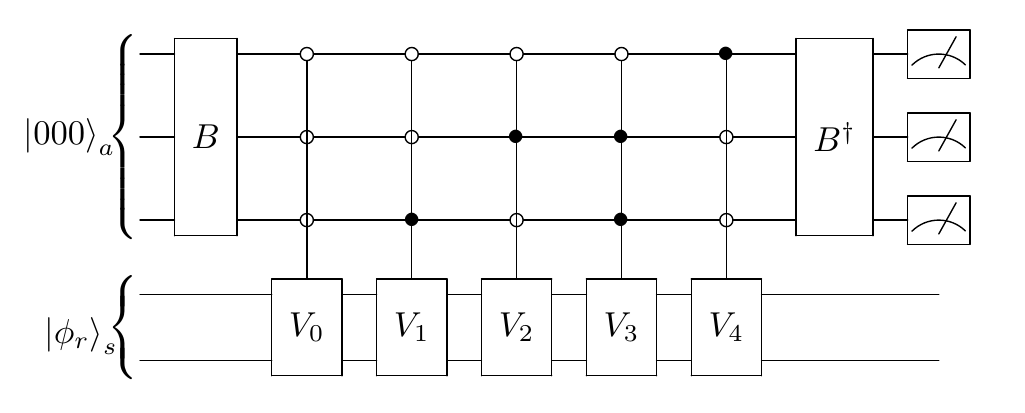}
    \caption{The quantum circuit to run Direct Measurement method when $n=2$. $B$ gate is prepared by the coefficients [1.31556, 0.13333, 0.13333, 0.25212, 1.06378]. $V_0$, $V_1$, $V_2$, $V_3$ and $V_4$ are applying $e^{-0.04180i}II$ and $e^{2.32888i}YY$,  $e^{2.32888i}XX$, $e^{3.05283i}ZI$ and $e^{3.11093i}IZ$ respectively.}
    \label{fig:circuit_2_v1}
\end{figure}

\begin{table}
    \centering
    \begin{tabular}{|c|c|c|}
    \hline
        Method  & Eigenenergy (Hartree) & Error (Hartree)\\
    \hline
    \hline
        Direct Diagonalization &  2.1259-0.1089i & -\\
    \hline
        Statevector Simulator & 2.1259-0.1089i & 0\\
    \hline
        QASM Simulator & 2.1279-0.1100i & $2 \times 10^{-3}$\\
    \hline
    \end{tabular}
    \caption{The complex eigenenergy obtained by directly diagonalizing the Hamiltonian and by running different simulators. The QASM simulator is configured to have no noise, and it takes $10^5$ samples to calculate the complex eigenenergy.}
    \label{tab:circuit_2_v1}
\end{table}

C2 in TABLE. \ref{tab:qubits_gates} is the case when we follow the steps for $n=2$ in section II and III. The Hamiltonian $H_\theta$ and how to calculate its complex eigenvalue are shown in Appendix D.A Eq.(\ref{eq:H_2}). FIG. \ref{fig:circuit_2_v1} gives the quantum circuit for $H_\theta$. This circuit can be executed in simulators with results listed in TABLE \ref{tab:circuit_2_v1}.\\

\begin{figure}[hbt!]
    \centering
    \includegraphics{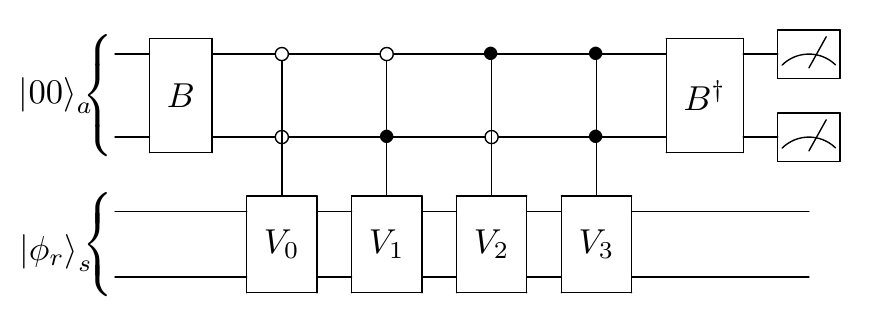}
    \caption{The simplified quantum circuit to run Direct Measurement method when $n=2$. $B$ gate is prepared by the coefficients [0.13333, 0.13333, 0.25212, 1.06378]. $V_0$, $V_1$, $V_2$ and $V_3$ are applying $e^{2.32888i}YY$,  $e^{2.32888i}XX$, $e^{3.05283i}ZI$ and $e^{3.11093i}IZ$ respectively.}
    \label{fig:circuit_2_v2}
\end{figure}

However, it is too complicated to be successfully run in IBM quantum computers. For C3 in TABLE \ref{tab:qubits_gates}, we simplify the quantum circuit by calculating the complex eigenvalue for the Hamiltonian $H_\theta$ in Appendix D.B, Eq. (\ref{eq:H_2_v2}). Because there are only 4 terms left, 2 ancilla qubits are enough for the method. The simplified quantum circuit is then shown in FIG. \ref{fig:circuit_2_v2}.
To avoid introducing more ancilla qubits, instead of $H'_\theta = H_\theta + xII$, we can run a similar 4-qubit circuit for $H'_\theta = H_\theta + H_\theta^3$, which has the same terms of tensor products as $H_\theta$ with different coefficients. 
\begin{table}[hbt!]
    \centering
    \begin{tabular}{|c|c|c|}
    \hline
        Method  & Eigenenergy (Hartree) & Error (Hartree)\\
    \hline
    \hline
        Direct Diagonalization &  2.1259-0.1089i & -\\
    \hline
        Statevector Simulator & 2.1259-0.1089i & 0\\
    \hline
        QASM Simulator & 2.1264-0.1099i & $1 \times 10^{-3}$\\
    \hline
        IBM Quantum Computer & 2.0700-0.4890i & 0.3841\\
    \hline
    \end{tabular}
    \caption{The complex eigenenergy obtained by directly diagonalizing the Hamiltonian, running simulators and using IBM quantum computers. The QASM simulator is configured to be noiseless, and it takes $10^{5}$ samples to calculate the complex eigenenergy. The IBM quantum computer takes $2^{13}$ samples.}
    \label{tab:circuit_2_v2}
\end{table}
This circuit can be executed successfully in the simulators and the IBM quantum computers. However, it costs around 200 gates in the IBM quantum computers, leading to significant error. The resulting resonance eigenenergies and errors can be seen in TABLE. \ref{tab:circuit_2_v2}. 

\begin{figure}[H]
    \centering
    \includegraphics{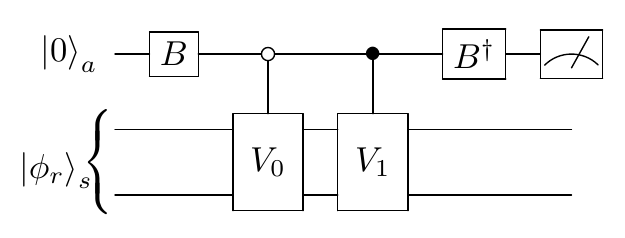}
    \caption{The quantum circuit to run Direct Measurement method when $n=2$. $B$ gate is prepared by the coefficients [1.19577, 0.53529]. $V_0$, $V_1$ are applying $e^{-0.09723i}II$ and $e^{-0.05311i}ZZ$ respectively.}
    \label{fig:circuit_2_v3}
\end{figure}

For the Hamiltonian Eq.(\ref{eq:H_2_v2}), a simpler circuit can be constructed if we try to calculate the complex eigenvalue of its square, Eq.(\ref{eq:H_2_v2_square}) in Appendix D.C. This is C4 in TABLE. \ref{tab:qubits_gates}. The quantum circuit for this $H_\theta^2$ is showed in FIG. \ref{fig:circuit_2_v3}.
\noindent
We can also run a similar 3-qubit circuit for $(H_\theta^2)'$ =  $H_\theta^2 + H_\theta^4$. The implementation of the circuit costs 9 gates in the IBM quantum computers after circuit optimization. The resulting eigenenergies are in TABLE. \ref{tab:circuit_2_v3}.

\begin{table}[H]
    \centering
    \begin{tabular}{|c|c|c|}
    \hline
        Method  & Eigenenergy (Hartree) & Error (Hartree)\\
    \hline
    \hline
        Direct Diagonalization &  2.1259-0.1089i & -\\
    \hline
        Statevector Simulator & 2.1259-0.1089i & 0\\
    \hline
        QASM Simulator & 2.1259-0.1107i & $1.7 \times 10^{-3}$\\
    \hline
        IBM Quantum Computer & 2.1624-0.1188i & 0.0378\\
    \hline
    \end{tabular}
    \caption{The complex eigenenergy obtained by directly diagonalizing the Hamiltonian, running simulators and running IBM quantum computers. The QASM simulator is configured to be noiseless, and it takes $10^{5}$ samples to calculate the complex eigenenergy. The IBM quantum computer takes $2^{13}$ samples. The error of the IBM quantum computer is from the best case.}
    \label{tab:circuit_2_v3}
\end{table}

\section{Quantum simulation of the  resonances in  H$_2^{-}$} 
This section presents a proof of concept that by using our quantum algorithm, the Direct Measurement method, one can calculate molecular resonances on a quantum computer. We focus on the resonances of a simple diatomic molecule,  $H_2^-$ $^2\Sigma_u^+(\sigma_g^2\sigma_u)$. Moiseyev and Corcoran  \cite{moiseyev1979autoionizing} showed how to obtain this molecule's resonance using a variational method based on the (5s,3p,1d/3s,2p,1d) contracted Gaussian atomic basis, which contains a total of 76 atomic orbitals for $H_2^-$. They picked around 45 configurations of natural orbitals as a final basis for resonance calculation. Here, however, we are not going to use their contracted Gaussian atomic basis, that needs 76 system qubits with additional ancilla qubits, which is too large to be simulated by classical computers. The number of gates would also be overwhelming. One may try an iterative diagonalization approach to get a few eigenvalues without constructing matrices or vectors. Another possible solution could be using tensor network simulators. Recent studies by Ellerbrock and Martinez show that tensor network simulators are able to efficiently  and accurately simulate over 100-qubit circuits with moderate entanglement \cite{ellerbrock2020multilayer}. In another study Zhou et al. showed that even strongly entangled systems (as those generated by 2D random circuits) can be simulated by matrix product states comparably accurate to modern quantum devices \cite{zhou2020limits}. However, building those simulators for our system is beyond the scope of this paper. In this way, we picked small basis sets, 6-31g, and cc-pVDZ, for our simulations. We used the Born-Oppenheimer approximation followed by complex rotation, as shown in Section II, "COMPLEX SCALED HAMILTONIAN" and mapped the Hamiltonian to the qubit space, as shown in Appendix  B. We then apply the Direct Measurement method to the Hamiltonian to obtain complex eigenvalues. An example quantum circuit to run the Direct Measurement method can be found in Appendix E.

FIG. \ref{fig:h2minus_figure} shows one complex eigenvalue's $\theta$ trajectories at $\alpha=1.00$ under different basis sets after running the algorithm. FIG. \ref{fig:h2minus_figure}.(a) is simulated using the 6-31g basis set. 8 spin-orbitals are considered in our self-defined simulator, and 16 qubits are needed to run the algorithm. In this case, if we fix $\eta=\alpha e^{-i\theta}$ at the lowest point in the figure, which has $\alpha=1.00$, $\theta=0.18$, the resonance energy obtained by the Direct Measurement method is $E_\theta = -0.995102 - 0.046236i$ Hartree. This complex energy is close to the one obtained in \cite{moiseyev1979autoionizing}, $E_\theta = -1.0995 -0.0432 i$ Hartree, especially the imaginary part. FIG. \ref{fig:h2minus_figure}.(b) is simulated using the cc-pVDZ basis set. We only considered the $s$ and $p_z$ basis functions for H atom for easier simulation. 12 spin-orbitals are considered in our self-defined simulator, and a total of 23 qubits are needed to run the algorithm in quantum computers. The results show the resonance energy at the lowest point in the figure, which has $\alpha=1.00$, $\theta=0.22$, is $E_\theta = -1.045083-0.044513 i$ Hartree. This is even closer to the one obtained in \cite{moiseyev1979autoionizing}. However, we want to note that the lowest points in FIG. \ref{fig:h2minus_figure} (a) and FIG. \ref{fig:h2minus_figure} (b) are not pause points. And they do not reveal real resonance properties. Also, even after shifting different $\alpha$ in simulations, we cannot find a consistent pause point in $\theta$ trajectories to locate the best resonance estimation. The reason may be related to our selected basis. Compared with the literature \cite{moiseyev1979autoionizing}, our basis set is much smaller, and is not optimized for the resonance state. Still, this application gives proof of concept and shows that one can calculate molecular resonances on a quantum computer. In the future, if more qubits are available in quantum computers, a large basis can be used, and we may be able to show finer structures in trajectories that can locate the best resonance point. Also, a larger basis set should lead to a more accurate resonance calculation.
\begin{figure}
    \begin{minipage}[t]{0.495\linewidth}
        \centering
        \includegraphics[width=\linewidth]{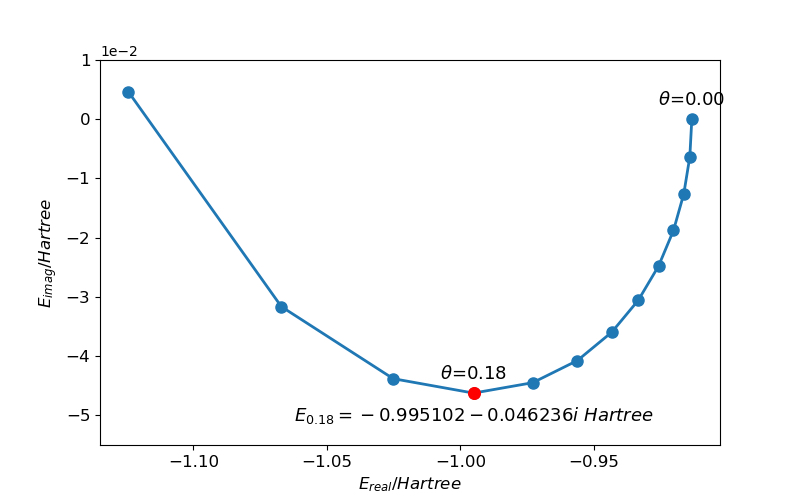}
        \subcaption*{\quad \quad (a)}
        \label{fig:h2minus:a}
    \end{minipage}
    \begin{minipage}[t]{0.495\linewidth}
        \centering
        \includegraphics[width=\linewidth]{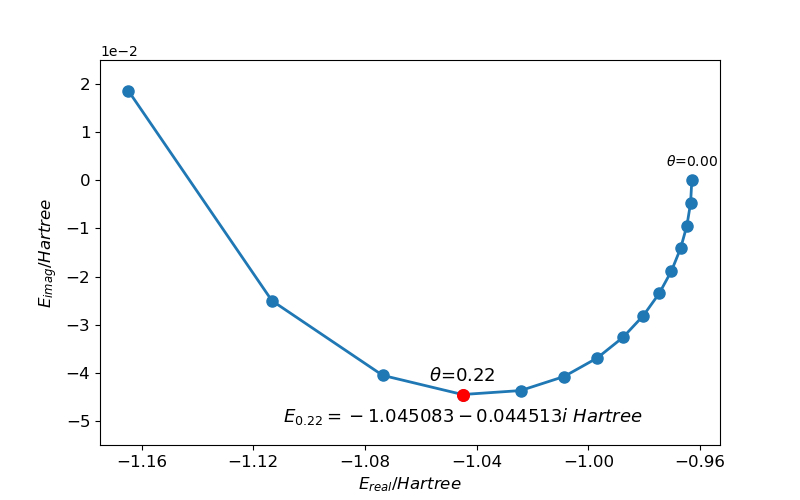}
        \subcaption*{\quad \quad (b)}
        \label{fig:h2minus:b}
    \end{minipage}
    \captionsetup{justification=raggedright,singlelinecheck=false}
    \caption{Complex eigenvalue trajectories on the rotation angle $\theta$ at $\alpha=1.00$ for molecule H$_2^-$, calculated by a self-defined simulator. (a) uses 6-31g basis set for H atom, including 1s and 2s orbitals. $\theta$ ranges from 0.00 to 0.24 with a step of 0.02.  At the lowest point when $\theta = 0.18$, the complex eigenvalue is $-0.995102 - 0.046236i$ Hartree. (b) uses the s and p$_z$ orbitals in cc-pVDZ basis set for H atom. $\theta$ ranges from 0.00 to 0.28 with a step of 0.02. At the lowest point when $\theta = 0.22$, the complex eigenvalue is $-1.045083 - 0.044513i$ Hartree.}
    \label{fig:h2minus_figure}
\end{figure}

\section{Conclusion}
In this paper, we construct and show a proof-of-concept for a quantum algorithm that calculates atomic and molecular resonances. We first presented the complex-scaling method to calculate molecular resonances.  Then, we introduced the Direct Measurement method, which embeds a molecular system's complex-rotated Hamiltonian into the quantum circuit and calculates the resonance energy and lifetime from the measurement results. These results represent the first applications of complex scaling Hamiltonian to molecular resonances on a quantum computer.  The method is proven to be accurate when applied to a simple one-dimensional quantum system that exhibits shape resonances. We tested our algorithm on quantum simulators and IBM quantum computers.  Furthermore, when compared to the exponential time complexity in traditional matrix-vector multiplication calculations, this method only requires $O(n^5)$ standard gates, where $n$ is the size of the basis set. These findings show this method's potential to be used in a more complicated molecular system and for better accuracy in the future when more and better qubit machines are available.

\section{Acknowledgements}
We would like to thank Rongxin Xia, Zixuan Hu, and Manas Sajjan for useful discussions. We also like to acknowledge the financial support by the National Science Foundation under award number  1955907.

\section{Data Availability}
The data that support the findings of this study are available from the corresponding author upon reasonable request.

\bibliography{ref.bib}

\newpage
\section*{Appendix A\\
\vspace{2em}
Complex-Rotated Hamiltonian of the Model System at $\theta=0.16$, $\alpha = 0.65$ when $n = 5$ }
\begin{center}
\begin{tabular}{||c | c||c | c|| c | c||}
\hline
 YYIII & -0.091665+0.096819i &  XXIII & -0.091665+0.096819i & IIIII & 4.599205-0.533073i \\
 ZIIII & -0.251131+0.022353i & YZYII &  0.0179156-0.030997i & XZXII &  0.0179156-0.030997i \\
 YZZYI & -0.007005+0.015446i & XZZXI & -0.007005+0.015446i & YZZZY  & 0.003680-0.009152i \\
 XZZZX & 0.003680-0.009152i & IZIII & -1.063280+0.032614i & IYYII &  -0.089297+0.108259i \\
 IXXII & -0.089297+0.108259i & IYZYI & 0.014213-0.055870i & IXZXI & 0.014213-0.055870i \\
 IYZZY & -0.003869+0.033693i & IXZZX & -0.003869+0.033693i & IIZII &  -1.445349+0.113618i \\
 IIYYI & -0.209952+0.010748i & IIXXI &  -0.209952+0.010748i &  IIYZY &  0.060302-0.008776j\\
 IIXZX & 0.060302-0.008776i & IIIZI &  -1.127058+0.243702i & IIIYY & -0.336956+0.051691i \\
 IIIXX & -0.336956+0.051691i & IIIIZ &  -0.712385+0.120784i & & \\
\hline
\end{tabular}
\captionof{table}{The coefficients and tensor product operators of the model system's complex-rotated Hamiltonian $H_\theta$ at $\theta = 0.16$, $\alpha = 0.65$ when there is $n = 5$ basis functions.}
\end{center}

\newpage
\section*{Appendix B\\
\vspace{2em}
Complex-Rotated Hamiltonian of H$_2^-$ at $\theta = 0.18$, $\alpha = 1.00$}
\begin{center}
\begin{tabular}{||c | c||c | c|| c | c||}
\hline
IXZXXZXI & 0.018705 -0.003404i & IIIZIXZX & 0.038191 -0.006950i & ZIZIIIII & 0.103932 -0.018913i  \\
XZXIXZXI & 0.027826 -0.005063i & IXXIIIXX & -0.002794+0.000508i & IIZZIIII & 0.106657 -0.019408i  \\
IYIYIIII & 0.024307 -0.004423i & IIIIXXXX & 0.015119 -0.002751i & IZIIIIZI & 0.095226 -0.017328i  \\
IIIIIIXX & 0.047512 -0.039979i & YYIIYZZY & -0.019254+0.003504i & XZXIYZYI & 0.027826 -0.005063i  \\
IZIIYZYI & 0.013080 -0.002380i & IIYYIIXX & 0.034554 -0.006288i & XZXIIIZI & 0.032587 -0.005930i  \\
YYYYIIII & 0.015119 -0.002751i & XXIIIYYI & 0.005216 -0.000949i & IXIXIIII & 0.024307 -0.004423i  \\
IIIIXXYY & 0.002918 -0.000531i & IIIZXZXI & 0.050249 -0.009144i & IIXXXXII & 0.020481 -0.003727i  \\
YYIIYYII & 0.019597 -0.003566i & IXXIIXXI & 0.008283 -0.001507i & IIIIXIXI & 0.016733 -0.003045i  \\
IYZYIIII & -0.035671+0.030324i & IYZYIIIZ & 0.043018 -0.007828i & YYIIIYYI & 0.005216 -0.000949i  \\
IIIIXZZX & -0.028316+0.033738i & XXIIYYII & 0.019597 -0.003566i & IXXIIIYY & -0.002794+0.000508i  \\
ZYZYIIII & 0.015436 -0.002809i & XXIIYZZY & -0.019254+0.003504i & IIIIIZZI & 0.084620 -0.015398i  \\
YZYIIZII & 0.011702 -0.002129i & IIYYXZZX & -0.031698+0.005768i & IIIIIXXI & -0.007550+0.006494i  \\
IXZXIZII & 0.012371 -0.002251i & IIIIYYYY & 0.015119 -0.002751i & IIIZYZYI & 0.050249 -0.009144i  \\
ZIIIIIII & -0.230405+0.108639i & ZIIIIIIZ & 0.159054 -0.028943i & IXXIYZZY & 0.006593 -0.001200i  \\
IIIIIYIY & 0.023153 -0.004213i & IIYYIXXI & -0.000541+0.000098i & YZZYIIII & -0.027204+0.031862i  \\
IIIZIIZI & 0.139579 -0.025399i & YZZYXXII & -0.016647+0.003029i & IIXXIIII & 0.047746 -0.040370i  \\
XIXIIIII & 0.017118 -0.003115i & YYIIXXII & 0.019597 -0.003566i & YZYIIYZY & 0.017127 -0.003117i  \\
IIIIZZII & 0.084496 -0.015376i & YZZYXZZX & 0.031161 -0.005670i & IIZIIYZY & 0.024717 -0.004498i  \\
XZZXIXXI & 0.004990 -0.000908i & IYYIIYYI & 0.008283 -0.001507i & IYZYIXZX & 0.015728 -0.002862i  \\
XZZXXZZX & 0.031161 -0.005670i & IYZYIIZI & 0.026040 -0.004739i & IIZIIZII & 0.093507 -0.017015i  \\
IZIIZIII & 0.106161 -0.019318i & XXIIIXXI & 0.005216 -0.000949i & IXZXIYZY & 0.015728 -0.002862i  \\
ZIIIIXZX & 0.030922 -0.005627i & IIIIIIYY & 0.047512 -0.039979i & XXIIIIYY & 0.021209 -0.003859i  \\
XXIIIIXX & 0.021209 -0.003859i & YYIIIIII & 0.001646 -0.022572i & ZIIIIIZI & 0.130169 -0.023687i  \\
IIYYIIYY & 0.034554 -0.006288i & YZYIIIIZ & 0.052229 -0.009504i & YZYIIIII & -0.021561+0.077956i  \\
IIXXXZZX & -0.031698+0.005768i & IIIIZYZY & 0.013729 -0.002498i & IYYIXXII & 0.003919 -0.000713i  \\
IIZIZIII & 0.133407 -0.024276i & YZZYYZZY & 0.031161 -0.005670i & XZXIZIII & 0.040337 -0.007340i  \\
ZIIIZIII & 0.151365 -0.027544i & YZYIIXZX & 0.017127 -0.003117i & IIIIYIYI & 0.016733 -0.003045i  \\
IIXXIYYI & -0.000541+0.000098i & IIYYYZZY & -0.031698+0.005768i & IYYIIIII & -0.009705+0.008779i  \\
YZZYYYII & -0.016647+0.003029i & XZXIIXZX & 0.017127 -0.003117i & IIIIIXIX & 0.023153 -0.004213i  \\
IIZIYZYI & 0.033580 -0.006110i & ZXZXIIII & 0.015436 -0.002809i & YZYZIIII & 0.020644 -0.003757i  \\
IIIIYZZY & -0.028316+0.033738i & IXZXZIII & 0.034152 -0.006215i & YZZYIYYI & 0.004990 -0.000908i  \\
ZIIZIIII & 0.126456 -0.023011i & YZZYIIXX & -0.029557+0.005379i & XZZXYZZY & 0.031161 -0.005670i  \\
IYYIIIYY & -0.002794+0.000508i & IXZXIIII & -0.035671+0.030324i & IXZXIIIZ & 0.043018 -0.007828i  \\
ZIIIYZYI & 0.038659 -0.007035i & IIXXIIXX & 0.034554 -0.006288i & ZZIIIIII & 0.085046 -0.015476i  \\
IIIZZIII & 0.158431 -0.028830i & YXXYIIII & 0.012162 -0.002213i & IZIIIYZY & 0.013159 -0.002395i  \\
IYZYXZXI & 0.018705 -0.003404i & XXIIXXII & 0.019597 -0.003566i & IIIIYXXY & 0.012201 -0.002220i  \\
IIIIZIII & -0.231557+0.112195i & IIIIZIIZ & 0.128680 -0.023416i & YZYIXZXI & 0.027826 -0.005063i  \\
IIYYYYII & 0.020481 -0.003727i & IIIIXZXZ & 0.020604 -0.003749i & IIIIXZXI & -0.030067+0.081498i  \\
IIYYIYYI & -0.000541+0.000098i & IYYIYZZY & 0.006593 -0.001200i & YZYIYZYI & 0.027826 -0.005063i  \\
IIZIXZXI & 0.033580 -0.006110i & IIXXYZZY & -0.031698+0.005768i & IIIIIIZI & -0.611815+0.267480i  \\
IIIIIIZZ & 0.107859 -0.019627i & YZZYIXXI & 0.004990 -0.000908i & IIIIIXZX & -0.012982+0.018373i  \\
\hline
\end{tabular}
\captionof{table}{The coefficients and tensor product operators in H$_2^-$'s complex-rotated Hamiltonian at $\theta = 0.18$, $\alpha = 1.00$ when using 6-31g basis set.}
\end{center}

\newpage
\section*{
\vspace{2em}
Complex-scaled Hamiltonian of H$_2^-$ at $\theta = 0.18$, $\alpha = 1.00$}

\vspace{-3em}
\begin{center}
    (Continued)
\end{center}

\begin{center}
\begin{tabular}{||c | c||c | c|| c | c||}
\hline
IXXIXXII & 0.003919 -0.000713i & IIZIIIII & -0.612966+0.271036i & XZXIIYZY & 0.017127 -0.003117i  \\
IIXXIXXI & -0.000541+0.000098i & IIIIYYII & 0.000598 -0.021276i & YYIIIIXX & 0.021209 -0.003859i  \\
XZZXYYII & -0.016647+0.003029i & XZXZIIII & 0.020644 -0.003757i & YZZYIIYY & -0.029557+0.005379i  \\
YYXXIIII & 0.002957 -0.000538i & YZYIIIZI & 0.032587 -0.005930i & IIXXYYII & 0.020481 -0.003727i  \\
IXZXIXZX & 0.015728 -0.002862i & IXZXIIZI & 0.026040 -0.004739i & XYYXIIII & 0.012162 -0.002213i  \\
ZIIIXZXI & 0.038659 -0.007035i & IIXXIIYY & 0.034554 -0.006288i & YYIIIIYY & 0.021209 -0.003859i  \\
IZZIIIII & 0.087497 -0.015922i & IZIIIZII & 0.094105 -0.017124i & IIYYXXII & 0.020481 -0.003727i  \\
IIIZIYZY & 0.038191 -0.006950i & IYYIIIXX & -0.002794+0.000508i & IXXIXZZX & 0.006593 -0.001200i  \\
IIIIZXZX & 0.013729 -0.002498i & IIIIIYYI & -0.007550+0.006494i & IIIIZIZI & 0.103932 -0.018913i  \\
YYIIXZZX & -0.019254+0.003504i & IXXIIIII & -0.009705+0.008779i & IIIIXXII & 0.000598 -0.021276i  \\
XZZXIYYI & 0.004990 -0.000908i & IZIIIIIZ & 0.110454 -0.020099i & IZIIIIII & -0.388873+0.102313i  \\
IYZYIZII & 0.012371 -0.002251i & IXXIIYYI & 0.008283 -0.001507i & IYYIYYII & 0.003919 -0.000713i  \\
YYIIIXXI & 0.005216 -0.000949i & XXYYIIII & 0.002957 -0.000538i & IXXIYYII & 0.003919 -0.000713i  \\
IIIIYZYZ & 0.020604 -0.003749i & IIIIYZYI & -0.030067+0.081498i & IYZYIYZY & 0.015728 -0.002862i  \\
IZIIXZXI & 0.013080 -0.002380i & IIIIIIII & 1.734311 -1.110499i & IIIIIIIZ & -0.896247+0.369556i  \\
IIZIIXZX & 0.024717 -0.004498i & IZIIIXZX & 0.013159 -0.002395i & IIZIIIZI & 0.120598 -0.021945i  \\
XZZXIIYY & -0.029557+0.005379i & IIIIXYYX & 0.012201 -0.002220i & IYZYZIII & 0.034152 -0.006215i  \\
IIYYIIII & 0.047746 -0.040370i & IXZXYZYI & 0.018705 -0.003404i & XZXIIIIZ & 0.052229 -0.009504i  \\
XZXIIIII & -0.021561+0.077956i & XZZXXXII & -0.016647+0.003029i & ZIIIIYZY & 0.030922 -0.005627i  \\
YIYIIIII & 0.017118 -0.003115i & IYYIXZZX & 0.006593 -0.001200i & XZZXIIII & -0.027204+0.031862i  \\
IIIIIYZY & -0.012982+0.018373i & XXIIXZZX & -0.019254+0.003504i & XZXIIZII & 0.011702 -0.002129i  \\
ZIIIIZII & 0.102700 -0.018688i & IIIIIZIZ & 0.092214 -0.016780i & IIIIIZII & -0.386698+0.100135i  \\
IYYIIXXI & 0.008283 -0.001507i & IIIZIZII & 0.105681 -0.019231i & XXIIIIII & 0.001646 -0.022572i  \\
IIIIYYXX & 0.002918 -0.000531i & IZIZIIII & 0.092214 -0.016780i & YZYIZIII & 0.040337 -0.007340i  \\
XXXXIIII & 0.015119 -0.002751i & XZZXIIXX & -0.029557+0.005379i & IIIZIIIZ & 0.184425 -0.033560i  \\
IIIZIIII & -0.894071+0.367379i & IIZIIIIZ & 0.144136 -0.026228i & IYZYYZYI & 0.018705 -0.003404i  \\
\hline
\end{tabular}
\captionof*{table}{TABLE V: The coefficients and tensor product operators in H$_2^-$'s complex-rotated Hamiltonian at $\theta = 0.18$, $\alpha = 1.00$ when using 6-31g basis set.}
\end{center}

\newpage
\section*{Appendix C}
\subsection*{How to get complex eigenvalue by the Direct Measurement Method}
\vspace{2em}
If the output state Eq. (\ref{eq:output_state}) is measured many times, the possibility of obtaining $\ket{0}_a$ state, $p$, is related to $E$ by the equation
\begin{align}
    p = \frac{E^2}{A^2},
    \label{eq:p1}
\end{align}
which reveals $|E| = \sqrt{p} A$. To obtain the phase, one way is that we apply a similar circuit for $H'_\theta = xI^{\otimes n} + H_\theta$, where $x$ is a selected real number. Then the updated $U_r'$ leads us to
\begin{align}
    p' = \frac{|x + Ee^{i\varphi}|^2}{A'^2}
    \label{eq:p2}
\end{align}
By applying $|E| = \sqrt{p} A$ to Eq. (\ref{eq:p2}), we can solve the phase $\varphi$ and finally the complex eigenvalue as
\begin{align}
    E e^{i\varphi} = \sqrt{p} A e^{i\cos^{-1}{\frac{p'A'^2 - x^2 - pA^2}{2xA\sqrt{p}}}} or \sqrt{p} A e^{-i\cos^{-1}{\frac{p'A'^2 - x^2 - pA^2}{2xA\sqrt{p}}}}.
    \label{eq:complex_energy}
\end{align}
If we expand the exponential term in Eq. (\ref{eq:complex_energy}), it becomes
\begin{align}
    E e^{i\varphi} = \frac{p'A'^2 - x^2 - pA^2}{2x} + i \frac{\sqrt{(2xA\sqrt{p})^2-(p'A'^2-x^2-pA^2)^2}}{2x}.
    \label{eq:complex_energy_alternative}
\end{align}
Since the measurement errors for $p$ and $p'$, i.e. $\Delta(p)$ and $\Delta{p'}$, are $O(\frac{1}{\sqrt{N}})$, based on Eq. (\ref{eq:complex_energy_alternative}) the error for the complex eigenvalue $E e^{i\varphi}$ is
\begin{align}
    \Delta(Ee^{i\varphi}) = O(\frac{1}{\sqrt{N}})
\end{align}
The larger the sampling size, the more accurate obtained complex eigenvalues are.\\

There are also other choices to obtain the phase. For example, instead of adding the $I^{\otimes n}$ part, we can try building the $U_r'$ based on $H_\theta + H_\theta^2$ or $H_\theta + H_\theta^3$ to get an equation like Eq.(\ref{eq:p2}) containing phase information. That equation together with Eq.(\ref{eq:p1}) will reveal the complex eigenvalue for the input eigenstate with another expression. 

\newpage
\section*{Appendix D}
\vspace{2em}
\subsection{The Hamiltonian and complex eigenvalue for the model system: n = 2, 5 qubits}
The complex-rotated Hamiltonian of the model system is
\begin{align}
\begin{split}
H_\theta = & 1.31556 * e^{-0.04180i}II + 0.13333 * e^{2.32888i}YY + 0.13333 * e^{2.32888i}XX\\
    & 0.25212 * e^{3.05283i}ZI + 1.06378 * e^{3.11093i}IZ.
\end{split}
\label{eq:H_2}
\end{align}

\noindent
By running the circuit FIG. \ref{fig:circuit_2_v1} for $H_\theta$ and a similar circuit for $H'_\theta = xII + H_\theta$, the complex eigenvalue can be derived by
\begin{align}
    E e^{i\varphi} = \sqrt{p} A e^{i\cos^{-1}{\frac{p'A'^2 - x^2 - pA^2}{2xA\sqrt{p}}}} or \sqrt{p} A e^{-i\cos^{-1}{\frac{p'A'^2 - x^2 - pA^2}{2xA\sqrt{p}}}},
\end{align}
where $A$ and $A'$ can be obtained from the absolute value of coefficients in $H_\theta$ and $H'_\theta$, $p$ and $p'$ can be obtained from measurement results.

\subsection{The Hamiltonian and complex eigenvalue for the model system: n = 2, 4 qubits}
The complex-rotated Hamiltonian of the model system without $II$ term is
\begin{align}
    H_\theta = &0.13333 * e^{2.32888i}YY + 0.13333 * e^{2.32888i}XX + 0.25212 * e^{3.05283i}ZI + 1.06378 * e^{3.11093i}IZ.
    \label{eq:H_2_v2}
\end{align}
\noindent
If we choose $H'_\theta = H_\theta + H_\theta^3$, which has the same terms of tensor products as $H_\theta$ with different coefficients, by running FIG.\ref{fig:circuit_2_v2} the complex eigenvalue for the original Hamiltonian can be represented by
\begin{align}
    Ee^{i\varphi} = &(1.31441-0.05497i) + \sqrt{p}A e^{\frac{i}{2} \cos^{-1}(\frac{p'A'^2}{2p^2A^4} - \frac{1}{2pA^2} - \frac{pA^2}{2})} \text{\ \ or}\\
    &(1.31441-0.05497i) + \sqrt{p}A e^{\frac{-i}{2} \cos^{-1}(\frac{p'A'^2}{2p^2A^4} - \frac{1}{2pA^2} - \frac{pA^2}{2})},
\end{align}
where $A$ and $A'$ can be obtained from the absolute value of coefficients in $H_\theta$ and $H'_\theta$, $p$ and $p'$ can be obtained from measurement results.

\subsection{The Hamiltonian and complex eigenvalue for the model system: n = 2, 3 qubits}

The square of the Hamiltonian Eq. (\ref{eq:H_2_v2}) is
\begin{align}
    H_\theta^2 = &1.19577*e^{-0.09723i}II + 0.53529 * e^{-0.05311i}ZZ
    \label{eq:H_2_v2_square}
\end{align}
If we choose $(H_\theta^2)' = H_\theta^2 + H_\theta^4$, by running FIG.\ref{fig:circuit_2_v3} the complex eigenvalue for the original Hamiltonian is 
\begin{align}
    Ee^{i\varphi} = &(1.31441-0.05497i) + p^{\frac{1}{4}}\sqrt{A} e^{\frac{i}{2} \cos^{-1}(\frac{p'A'^2}{2p^{\frac{3}{2}}A^3} - \frac{1}{2\sqrt{p}A} - \frac{\sqrt{p}A}{2})} \text{\ \ or}\\
    &(1.31441-0.05497i) + p^{\frac{1}{4}}\sqrt{A} e^{\frac{-i}{2} \cos^{-1}(\frac{p'A'^2}{2p^{\frac{3}{2}}A^3} - \frac{1}{2\sqrt{p}A} - \frac{\sqrt{p}A}{2})}
\end{align}
where $A$ and $A'$ can be obtained from the absolute value of coefficients in $H_\theta^2$ and $H_\theta^2 + H_\theta^4$, $p$ and $p'$ can be obtained from their measurement results.

\newpage
\section*{Appendix E\\
\vspace{2em}
Quantum Circuit for Complex-scaled Hamiltonian of H$_2^-$ at $\theta = 0.18$, $\alpha = 1.00$}

The complex-scaled Hamiltonian of $H_2^-$ at $\theta = 0.18$, $\alpha = 1.00$ in Appendix B can be written as
\begin{align}
\begin{split}
    H = 
     & 0.019012*e^{-0.180013i} IXZXXZXI + 0.038818*e^{-0.180010i} IIIZIXZX + \\
     & 0.105638*e^{-0.180005i} ZIZIIIII + 0.028282*e^{-0.179983i} XZXIXZXI + \\
     & ... + 0.019012*e^{-0.180013i} IYZYYZYI.
\end{split}
\label{eq:ap_h2minus_ham}
\end{align}
We would like to mention that the terms explicitly shown in Eq. (\ref{eq:ap_h2minus_ham}) are following the order in Appendix B. It is a coincident that their phases are similar. For example, one term we didn't show in the Hamiltonian is $0.021284*e^{1.542696i}IIIIYYII$, which has a different phase.\\

\begin{figure}
    \centering
    \includegraphics{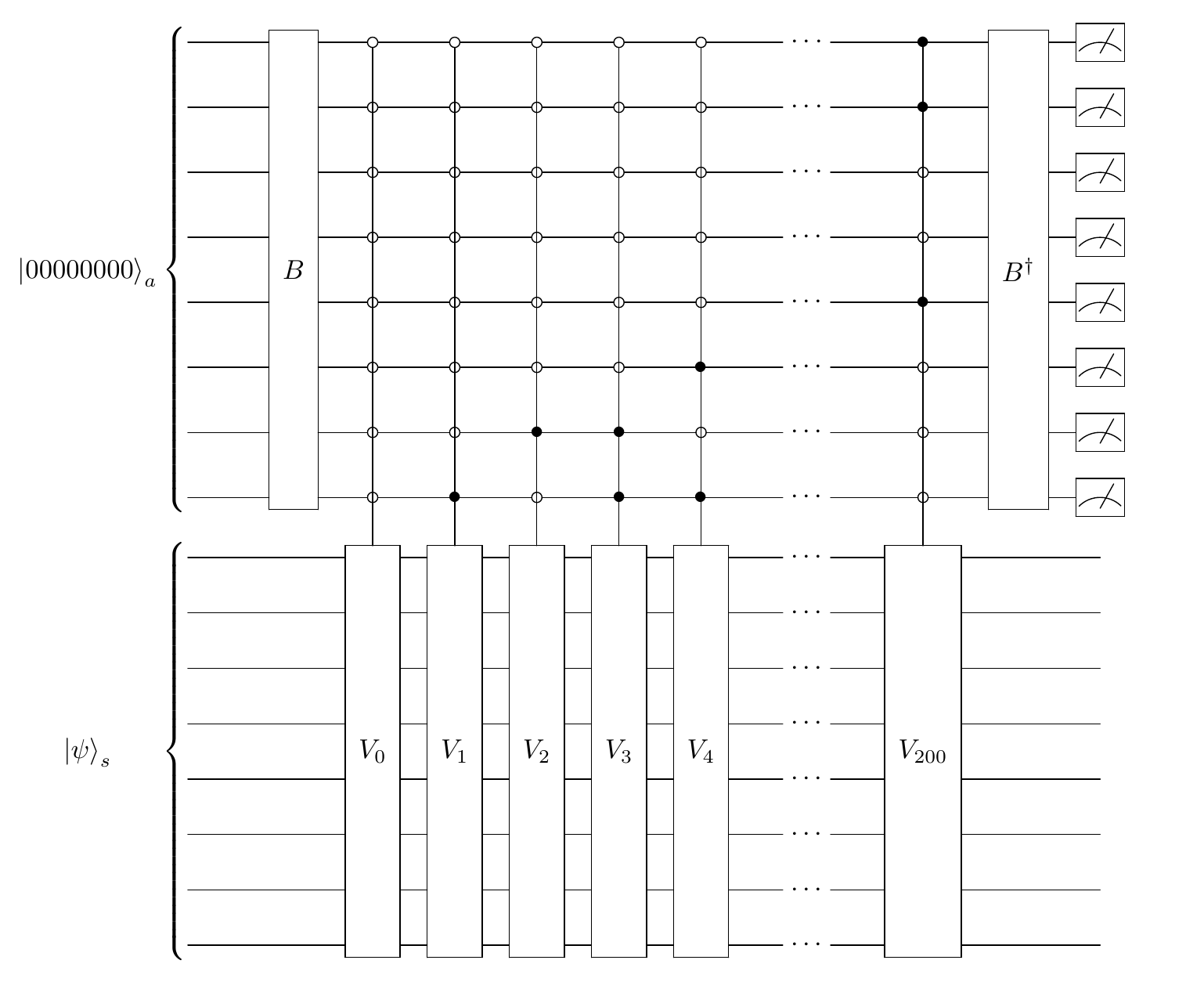}
    \captionsetup{justification=raggedright,singlelinecheck=false}
    \caption{The quantum circuit to run the Direct Measurement method for H$_2^-$ when $\theta = 0.18, \alpha=1.00$.  $B$ gate can be prepared by $\boldsymbol{\beta}$ in Eq. (\ref{eq:h2minus_ham_coeff}). $V_i$ gates are listed in Eq. (\ref{eq:h2minus_ham_operator}).}
    \label{fig:qcircuit_h2minus}
\end{figure}

To construct the quantum circuit for Direct Measurement method, we need to create $B$ gate and $V$ gate. The $B$ gate can be prepared by the coefficients from the Hamiltonian Eq. (\ref{eq:ap_h2minus_ham})
\begin{align}
    \boldsymbol{\beta} =
    \kbordermatrix{
        index &\\
        \vspace{1.5em}
        0 & 0.019012\\
        \vspace{1.5em}
        1 & 0.038818\\
        \vspace{1.5em}
        2 & 0.105638\\
        \vspace{1.5em}
        3 & 0.028282\\
        \vspace{1.5em}
        \vdots & \vdots\\
        \vspace{1.5em}
        200 & 0.019012\\
        \vspace{1.5em}
        201 & 0\\
        \vspace{1.5em}
        202 & 0\\
        \vspace{1.5em}
        \vdots & \vdots\\
        255 & 0
    },
\label{eq:h2minus_ham_coeff}
\end{align}
as shown in Eq. (\ref{eq:bgate_construction}). The $V$ gate can be constructed by a series of controlled-$V_i$ gates, where $V_i$ are
\begin{align}
    &V_0 = e^{-0.180013i} IXZXXZXI,\\
    &V_1 = e^{-0.180010i} IIIZIXZX,\\
    &V_2 = e^{-0.180005i} ZIZIIIII,\\
    &V_3 = e^{-0.179983i} XZXIXZXI,\\
    &\vdots\\
    &V_{200} = e^{-0.180013i} IYZYYZYI.
\label{eq:h2minus_ham_operator}
\end{align}
The whole circuit is shown in the FIG.  \ref{fig:qcircuit_h2minus}. The encoding of control qubits is based on the binary form of $V_i$'s index $i$. For example,  $V_3$ is applied to $\ket{\psi}_s$ if the ancilla qubit state is $\ket{3}_a = \ket{00000011}_a$.

\end{document}